\documentclass[english,twocolumn,aps,prl,showpacs]{revtex4-1}

\usepackage[T1]{fontenc}
\usepackage[latin1]{inputenc}
\usepackage{graphicx}
\usepackage{amssymb}
\usepackage{amsfonts}
\usepackage{amsmath}
\usepackage{stmaryrd}

\DeclareMathAlphabet{\mathsfsl}{OT1}{cmss}{m}{sl}



\usepackage{babel}
\makeatother

\begin{document}

\title{Criteria for 2D kinematics in an interacting Fermi gas}

\author{P. Dyke$^1$, K. Fenech$^1$, T. Peppler$^1$, M. G. Lingham$^1$, S. Hoinka$^1$, W. Zhang$^2$, S.-G. Peng$^3$, B. Mulkerin$^1$, H. Hu$^1$, X.-J. Liu$^1$, and C. J. Vale$^{1*}$}
\affiliation{$^1$Centre for Quantum and Optical Sciences, Swinburne University of Technology, Melbourne 3122, Australia. \\ 
$^2$Department of Physics and Beijing Key Laboratory of Opto-electronic Functional Materials and Micro-nano Devices, Renmin University of China, Beijing 100872, China. \\ 
$^3$State Key Laboratory of of Magnetic Resonance and Atomic and Molecular Physics, Wuhan Institute of Physics and Mathematics, Chinese Academy of Sciences, Wuhan 430071, China. \\
{$^\ast$To whom correspondence should be addressed; E-mail:  cvale@swin.edu.au}}

\date{\today}

\begin{abstract}
Ultracold Fermi gases subject to tight transverse confinement offer a highly controllable setting to study the two-dimensional (2D) BCS to Berezinskii-Kosterlitz-Thouless superfluid crossover. Achieving the 2D regime requires confining particles to their transverse ground state which presents challenges in interacting systems. Here, we establish the conditions for an interacting Fermi gas to behave kinematically 2D. Transverse excitations are detected by measuring the transverse expansion rate which displays a sudden increase when the atom number exceeds a critical value $N_{2D}$ signifying a density driven departure from 2D kinematics. For weak interactions $N_{2D}$ is set by the aspect ratio of the trap. Close to a Feshbach resonance, however, the stronger interactions reduce $N_{2D}$ and excitations appear at lower density.

\end{abstract}


\maketitle
Fermions confined to two-dimensional (2D) planes represent an important paradigm in many-body physics in settings ranging from thin films of superfluid helium-3 \cite{davis88,levitin13} to the superconducting planes in high-$T_c$ cuprates~\cite{tinkham04}. Ultracold atomic gases confined in oblate potentials allow access to the 2D regime~\cite{gorlitz01, modugno03, smith05, gunter05, stock05, du09, clade09, gillen09, martiyanov10, frohlich11, dyke11, ries14} where interactions between particles can be controlled using a Feshbach resonance~\cite{chin10}. In 2D Fermi gases, one can realize the BCS to Berezinskii-Kosterlitz-Thouless (BKT) superfluid crossover~\cite{berezinskii71,kosterlitz73, petrov03, bothelo06, zhang08a, iskin09, bertaina11, fischer13, bauer14} by tuning the attractive interaction between particles in different spin states. Of particular interest is the enhanced pairing due to the transverse confinement~\cite{petrov01,bloch08,feld11,sommer12, zhang12} and the consequences this has for the phase diagram of the crossover~\cite{zhang08b, makhalov14, ries14, levinsen14}.

Theoretical studies of the BCS-BKT crossover generally assume only two spatial dimensions, however, all atomic gases exist in 3D environments. Lower dimensional behaviour can be realized by freezing out dynamics along one or more directions. For atoms in a harmonic potential, with frequencies $\omega_x, \omega_y$ and $\omega_z$, the 2D regime is achieved when the transverse ($z$) confinement is strong enough that occupation of transverse excited states is energetically forbidden. When a gas is frozen in the transverse ground state, dynamics in the $x$-$y$ plane become decoupled from $z$ and the gas is kinematically 2D. In an ideal gas this requires the thermal energy and chemical potential be much smaller than the transverse confinement energy $k_B T, \mu \ll \hbar \omega_z$, where $k_B$ is Boltzmann's constant, $T$ is the temperature and $\mu$ the chemical potential. When interactions are present, however, these can provide another means for generating transverse excitations which go beyond purely 2D models.

In this Rapid Communication, we examine the criteria for an interacting Fermi gas to behave kinematically 2D. By measuring the transverse cloud width after time of flight we observe a rapid growth in the expansion rate when transverse excitations are present. Both the trap geometry and interaction strength are seen to limit the parameter space where interacting systems are kinematically 2D. An exact two-body calculation provides zero-density baseline to help understand our observations.

In an ideal Fermi gas Pauli exclusion sets an upper limit on the allowed atom number $N$ of a 2D system as $T \rightarrow 0$. For finite $\omega_z$ there can only be a finite number of single particle states with energy below the first transverse excited state. In a cylindrically symmetric trap ($\omega_x = \omega_y = \omega_r$) the critical number of states is given by $N_{2D}^{(Id.)} \approx (\omega_z / \omega_r)^2$~\cite{dyke11}. For $N > N_{2D}^{(Id.)}$ the Fermi energy $E_F$ will exceed $\hbar \omega_z$, the gas is no longer restricted to the transverse ground state and excited states play a visible role~\cite{schneider98,petrov01,vignolo03,mueller04,dyke11,hu11}.

A more complex scenario arises in an interacting gas. Neutral atoms interact via the van der Waals potential which has a range $r_0$ much smaller than the typical transverse quantization length, $\ell_z = \sqrt{\hbar /(m \omega_z)}$, where $m$ is the atomic mass, achievable in experiments. At length scales $r$, in the range $r_0 < r < \ell_z$, atomic scattering is barely modified by the transverse confinement and the relative wave function has the same form as in 3D, differing only in the normalization~\cite{petrov01}. Low energy 3D scattering is characterized by the $s$-wave scattering length $a$ which can be tuned using a Feshbach resonance. In 3D, stable two-body dimers only exist for $a > 0$. In quasi-2D, however, the transverse confinement gives rise to a two-body bound state for all $a \neq 0$ and the magnitude of the molecular binding energy $E_b$ is always greater than in the 3D case~\cite{petrov01,bloch08}. The size and character of the dimer state is set by $E_b$. At the two-body level, when $E_b \ll \hbar \omega_z$ the transverse confinement dominates and atoms remain primarily in the transverse ground state. When $E_b \gg \hbar \omega_z$ molecules are tightly bound with a size small compared to $\ell_z$ and are essentially identical to 3D molecules with equal atomic motion in all three dimensions. In the $T \rightarrow 0$ dilute limit even when $E_b <  \hbar \omega_z$ interactions significantly modify the ground state wavefunction~\cite{idziaszek05,kestner06}. At finite densities, the situation is less understood and transverse excitations can play a strong role in quasi-2D gases~\cite{fischer13,merloti13,fischer14,levinsen14}. 

    \begin{figure}[!t]
        \centering
        \includegraphics[clip,width=0.48\textwidth]{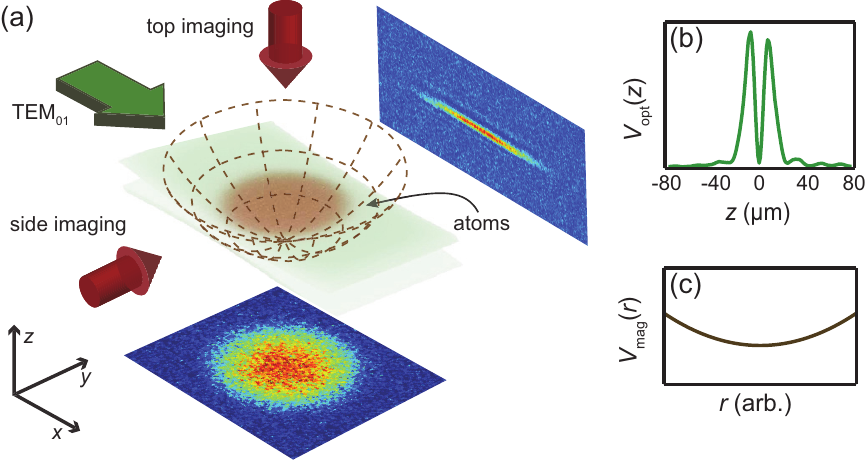}
        \caption{(a) Experimental setup for producing 2D clouds. Atoms are confined between the antinodes of a 532 nm blue-detuned TEM$_{01}$ mode laser beam. Harmonic radial confinement is produced by the residual curvature in the magnetic field generated by the Feshbach coils. (b) Transverse profile of the TEM$_{01}$ mode which defines $V_z$ and (c) harmonic radial trapping potential.}
        \label{fig:fig1}
    \end{figure}

Whether an interacting Fermi gas is kinematically two-dimensional therefore depends on the trap geometry, temperature, chemical potential and interactions. Here, we address this question using a highly degenerate Fermi gas of $^6$Li atoms confined in an oblate potential with interactions tuned via a broad Feshbach resonance at 832.2 G~\cite{zurn13}. Figure 1(a) shows the experimental setup. A 2D trap is formed between the antinodes of a cylindrically focused, 532 nm (blue-detuned), TEM$_{01}$ mode laser beam~\cite{smith05, rath10}, with $1/e^2$ radii of $\textit{w}_z$ $\approx$ 10 $\mu$m and $\textit{w}_x \approx 1.4$ mm, which produces the tight confinement in the $z$ direction and very weak anti-confinement in the $x$-$y$ plane. Residual magnetic field curvature from the Feshbach coils provides highly harmonic and cylindrically symmetric confinement in the $x$-$y$ plane which completely dominates the anti-trapping of the optical potential. The combined optical and magnetic potentials are plotted in Fig.~1(b) and (c), respectively. The measured trapping frequencies are $\omega_z/2\pi$ = 5.15 kHz and $\omega_r/2\pi$ = 26.4 Hz at a magnetic field of $B$ = 972 G. The trap aspect ratio here is approximately 200 which yields $N_{2D}^{(Id.)} = 3.9 \times 10^4$ atoms. Note that $\omega_r \propto \sqrt{B}$ while $\omega_z$ is independent of $B$ such that $N_{2D}^{(Id.)}$ increases as $B$ decreases.

To produce a 2D Fermi gas we begin with a 3D cloud of approximately $N = 4 \times 10^5$ $^6$Li atoms in each of the lowest two spin states $|F = 1/2, m_F = \pm 1/2\rangle$ evaporatively cooled in a 1075 nm single beam (3D) optical dipole trap. With the magnetic field at 780 G (BEC side of the Feshbach resonance) we ramp on the 2D optical potential in 350 ms and subsequently ramp off the single beam optical trap in 350 ms. The TEM$_{01}$ laser beam is initially turned on at low power, and $B$ is set to 832.2 G where the 3D scattering length diverges. A magnetic field gradient is then applied along $z$ over 2 seconds to achieve further evaporative cooling of the quasi-2D cloud which also allows the atom number to be precisely controlled. Next, we simultaneously ramp up the TEM$_{01}$ beam to full power and ramp down the magnetic field gradient in 750 ms. Finally, we sweep $B$ to the desired value in 250 ms and hold the cloud for 50 ms before taking an image either {\it in situ} or after time of flight. Absorption imaging is achieved by illuminating the cloud with a 10 $\mu$s pulse of resonant laser light at approximately half of the saturation intensity. We can image from the top to directly obtain the 2D density $n(x,y)$ or from the side time to obtain the transverse size of an expanded cloud, Fig.~1(a). We produce clouds with $N$ between $4 - 100 \times 10^3$ atoms per spin state. We estimate the temperature of the 2D clouds to be $0.1 \, T_F$ where $T_F = E_F/k_B$ is the Fermi temperature and $E_F = \sqrt{N} \hbar \omega_r$ in a harmonic trap, by fitting a 2D Thomas-Fermi profile to a weakly interacting cloud with $N = 24 \times 10^3 (= 0.6 N_{2D}^{(Id.)})$ at $B = 972$ G where $E_b/E_F \approx 0.001$. 

In a first set of experiments we measured the transverse cloud width $\langle z^2 \rangle^{1/2}$, at time $\tau$ = 600 $\mu$s after switching off the TEM$_{01}$ laser as the magnetic field was tuned across the Feshbach resonance. This expansion time is long enough to resolve the size along $z$ ($\tau \gg 1/\omega_z$) but short enough ($\tau \ll 1/\omega_r$) that no changes were detected in the radial density profile. Suddenly removing the transverse confinement leaves the cloud out of equilibrium with excess transverse kinetic energy that drives rapid expansion along $z$. As the magnetic field is kept on after release, the $s$-wave scattering length remains high and the gas expands in a collisional regime. While this prevents a direct extraction of the {\it in situ} transverse momentum distribution, the signature of transverse excitations is readily visible in the presence of collisions \cite{supplement}. 

    \begin{figure}[!t]
        \centering
        \includegraphics[clip,width=0.45\textwidth]{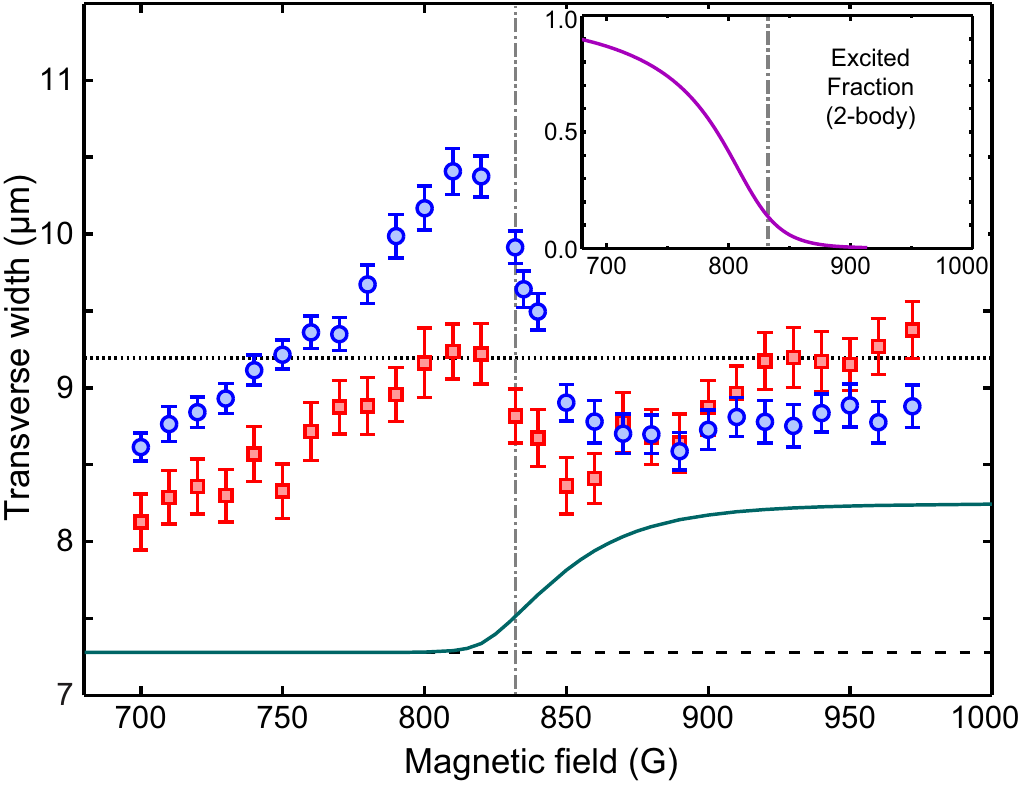}
        \caption{Transverse cloud width after 600 $\mu$s expansion versus magnetic field around the broad Feshbach resonance ($B_0 = 832.2$ G, grey dash-dotted line) for $N = 24 \times 10^3 ( = 0.6 N_{2D}^{(Id.)})$ (blue circles) and $N = 12 \times 10^3 (= 0.3 N_{2D}^{(Id.)}$) (red squares). Horizontal dotted (dashed) black line represents the expected width of a non-interacting atomic (molecular) cloud. Green solid line shows the expected width of an expanded cloud in the dilute limit at $T = 0$ based on a two-body calculation and the inset shows the transverse excited state fraction in the same limit.}
        	\label{fig:fig2}
    \end{figure}

In Fig.~2 we plot $\langle z^2 \rangle^{1/2}$ for $N = 0.6 N_{2D}^{(Id.)}$ which reveals a unique feature associated with a quasi-2D gas. The cloud width peaks on the BEC side of the Feshbach resonance, in stark contrast to the 3D case, where one observes a monotonic increase in the expanded cloud size when tuning from the BEC-BCS regimes~\cite{bourdel04}.

In the absence of interactions, all atoms in a gas with this atom number ($N <  N_{2D}^{(Id.)}$) and temperature would occupy the transverse ground state. The expanded cloud would have a Gaussian transverse profile with a width set by the momentum distribution of the zero-point motion in the trap. The dotted line in Fig.~2 indicates the expected width of the corresponding Gaussian wavepacket after expansion 9.2 $\mu$m, which includes the 4.7 $\mu$m (rms) resolution of our imaging system. Only when $N$ exceeds $N_{2D}^{(Id.)}$ and atoms are forced to fill higher lying states would we see the transverse width increase due to a broader momentum distribution. 

In the far BCS limit, where $a$ is negative and relatively small, $E_b \ll E_F$ and we approach this ideal gas behavior. For magnetic fields above 870 G the observed cloud width is approximately constant in good agreement with the ideal case. At lower fields, $820$ G $\lesssim B \lesssim$ 860 G, the transverse width begins increasing signifying a departure from 2D kinematics. The combination of strong interactions and finite density (which sets the relative collision energy via the local Fermi energy $\varepsilon_F = \pi \hbar^2 n /m$) couple energy into the transverse dimension.

On the BCS side of the Feshbach resonance, any pairs present in the quasi-2D trap will dissociate upon release and we image a cloud of expanding atoms. Below 832 G the interpretation of the data is slightly more complicated as stable dimers may exist in 3D and removing the confinement will not necessarily break molecules. Figure 2 shows that the transverse width continues to grow between 832 G to 820 G indicating that 3D molecules in this range tend to dissociate in expansion. This is consistent with the energy scales as the energy of the zero-point motion in the initial trap is large compared to the 3D binding energy ($E_{b,3D}/(\hbar \omega_z) = 0.1$ at 820 G). At even lower fields ($<$ 800 G), the final binding energy in 3D eventually exceeds $\hbar \omega_z$ and molecules remain bound after removing the confinement. Then we image a gas of repulsively interacting molecules. As $B$ is further reduced, the repulsion between molecules becomes weaker and the width after time of flight decreases towards the noninteracting molecule limit (7.3 $\mu$m for our imaging system) dashed line Fig.~2. When $E_b \gg \hbar \omega_z$, we recover the behavior for a quasi-2D Bose gas and the notion of $N_{2D}^{(Id.)}$ becomes irrelevant as Pauli exclusion no longer plays a role. In the radial direction (not shown) we observe the cloud width to shrink monotonically from the BCS to BEC regimes, similar to a 3D Fermi gas~\cite{bourdel04}. 

Also shown in Fig.~2 is the same measurement repeated at a lower atom number $N = 12 \times 10^3 \; (0.3 N_{2D}^{(Id.)})$ which displays a similar shape, but with a smaller and narrower peak around 810 G. At high fields the measured width lies slightly above the ideal gas prediction most likely due to the finite temperature and weaker interactions. On the BEC side of the Feshbach resonance the lower density reduces the effect of interactions so that the width at 700 G is closer to the noninteracting molecule prediction (7.3 $\mu$m). The differences between these two sets of data highlight the role of the atomic density. Higher densities enhance transverse excitations as seen by the taller and broader peak in the large $N$ data set. In both cases however, the peak lies at the same magnetic field, indicating that its location is set by the ratio $E_b / (\hbar \omega_z)$. 

Theoretically, we can gain insight into these observations by comparison with exact solutions for the two-body ground state wavefunction both {\it in situ} and after expansion. In a quasi-2D trap, a bound state always exists for arbitrarily weak attraction and the axial size of the wavefunction for the relative motion is smaller than the single particle ground state~\cite{idziaszek05,kestner06}. This can be quantified by projection onto the harmonic oscillator states which shows a significant excited atomic fraction that increases monotonically below 900 G approaching unity in the BEC limit when $E_b \gg \hbar \omega_z$~\cite{kestner06}, inset Fig.~2. 

We can also calculate the evolution of the two-body wavefunction following release from the trap and hence the time-dependence of $\langle z^2 \rangle^{1/2}$ \cite{supplement}. Convoluting this with our imaging resolution, we obtain a prediction of how the observed cloud width would scale in the dilute limit ($N \rightarrow 0$). This is shown as the solid green line in Fig.~2 which reveals a monotonic increase from the BEC to BCS regimes that lies below our measured widths. Even in the weakly attractive limit ($B \gtrsim 900$ G) the two-body result lies well below the ideal gas. This is because radial confinement remains during expansion, both in the experiment and calculation, and the ground state is still (extremely) weakly bound with $E_b / (2 \pi \hbar)$ of order a Hz. In the experiments however, thermal fluctuations will exceed this level leaving a gas of freely expanding atoms. The most notable difference between the two-body calculation and the data is the peak near 810 G in the experiments, which highlights that the combination of density and interactions, not simply the modified two-body wavefunction, contribute to the excitations.

In a complementary study we fixed the magnetic field and measured the transverse cloud width after time of flight as a function of atom number. As seen previously for a weakly interacting gas~\cite{dyke11}, one observes an elbow in the transverse width, at $N_{2D}$, when new transverse states become energetically accessible. In Fig.~3(a-c) we plot curves demonstrating such features at fields of 832 G, 865 G and 950 G. We determine $N_{2D}$ from the intersect of two straight line fits to the (filled) data points above and below the elbow. For the weakly interacting clouds (950 G), where the 2D binding energy $E_b =  0.0015  \hbar \omega_z$, one observes a clear elbow at $N \approx 40 \times 10^3$ consistent with the geometric ideal gas limit on the population of the transverse ground state $N_{2D}^{(Id.)}$. At 865 G ($E_b =  0.04 \, \hbar \omega_z$) we observe similar behavior with a clear plateau in the transverse width, however, the elbow is shifted to a lower atom number $(\approx 20 \times 10^3)$. Here the departure from 2D is driven by interactions as the peak Fermi energy at the trap center $\varepsilon_F = 0.83(9) \hbar \omega_z$ and $N_{2D}^{(Id.)} \approx 42 \times 10^3$ due to the weaker radial confinement. Closer to the Feshbach resonance the elbow shifts to even lower ratios of $\varepsilon_F / (\hbar \omega_z)$. Once we reach the Feshbach resonance at 832 G ($E_b = 0.244 \hbar \omega_z$) the plateau corresponding to the transverse ground state is no longer visible for the atom numbers we can access.

    \begin{figure}[!t]
        \centering
        \includegraphics[clip,width=0.48\textwidth]{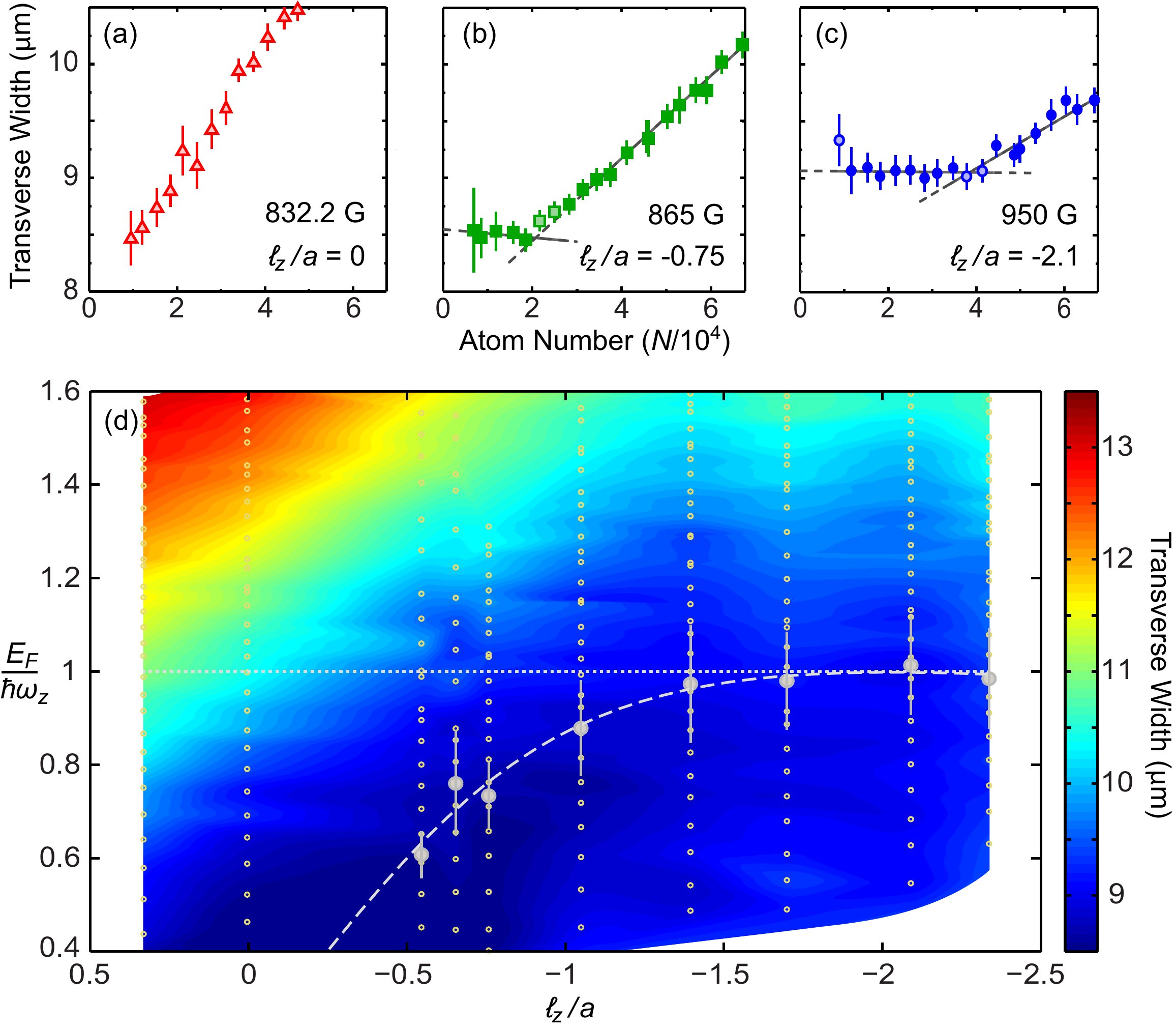}
        \caption{Transverse cloud width after time of flight versus atom number at magnetic fields of (a) 832 G, (b) 865 G, and (c) 950 G. Straight lines are fitted to the filled data points and the intersect is used to determine $N_{2D}$. (d) Transverse width (false color) as a function of the ratio of the 2D Fermi energy $E_F$ and the interactions expressed as $\ell_z/a$. Light grey circles indicate the location of the elbows showing the measured critical atom numbers at different magnetic fields. Small open circles indicate all measurement points. Light grey dashed line is a guide to the eye indicating the upper boundary the strictly 2D regime. Dotted white line indicates $N_{2D}^{(Id.)}$.}
        \label{fig:fig3}
    \end{figure}

We have performed a series of measurements like those in Fig.~3(a-c) for a range of magnetic fields to construct an image of the transverse cloud width as a function of atom density and interactions. This is shown in false color in Fig.~3(d) versus the 2D Fermi energy $E_F = \sqrt{N} \hbar \omega_r$ and interaction strength in units of $\ell_z / a$. For each set of width measurements (small open circles) we determined the critical atom number $N_{2D}$, as for Figs.~3(a-c), above which the cloud has clearly detectable transverse excitations (light grey circles). In the region below the dashed line the gas behaves kinematically 2D while above this, transverse excitations are present. This plot highlights that the gas is only kinematically 2D over a relatively small region with weak attractive interactions corresponding to the 2D BCS limit. When $-\ell_z/a$ drops below $~1.5$ (or $E_b/ (\hbar \omega_z) > 0.007$) interactions reduce the size of the 2D regime below the geometric limit. Even at $B = 860$ G and $N = 10^4$, just inside the 2D regime, we have $E_b / E_F \approx 0.1$ highlighting that modest interactions can produce transverse excitations at nonzero density. The geometric upper limit for 2D kinematics ($E_F = \hbar \omega_z$), indicated by the white dotted line only applies in the far BCS limit. For comparison, the data in Fig.~2 correspond to horizontal cuts though Fig.~3(d) at $E_F /(\hbar \omega_z) \approx$ 0.75 and 0.5 for $N = 0.6 N_{2D}^{(Id.)}$ and $N = 0.3 N_{2D}^{(Id.)}$, respectively. 

In summary, we have characterized the parameter range for which an interacting quasi-2D Fermi gas is kinematically 2D. Increasing the density enhances transverse excitations beyond two-body predictions as interactions become stronger. Both $E_b \ll E_F$ and $E_F, k_B T < \hbar \omega_z$ are necessary for the center of mass (bosonic) and relative (fermionic) motional degrees of freedom to remain 2D. While it is unclear whether the excitations we observe are atomic or molecular in character, we note that the slight anharmonicity in the transverse potential produced by TEM$_{01}$ mode laser may allow the formation of vibrationally excited molecules \cite{haller10,peng11,sala12}. Such anharmonicities will be present in nearly all low-dimensional atom traps and similar behavior could be expected in a broad range of experiments. Our findings provide a basis for when 2D theoretical treatments of interacting fermions can be applied to experiments on the BCS-BKT crossover. Studies of the equation of state of a 2D Fermi gas \cite{bertaina11,bauer14,makhalov14} will need to take account of transverse excitations in the strongly interacting regime.

We thank P. Hannaford, M. Parish and J. Levinsen for helpful discussions.  C. J. V. and P. D. acknowledge financial support from the Australian Research Council programs FT120100034, DP130101807 and DE140100647.

\makeatletter

\newpage

\makeatother

\makeatother


%
%
%

\onecolumngrid

\begin{centering} 
\textbf{\large Supplemental material: Critieria for 2D kinematics in an interacting Fermi gas} \\
\bigskip 

P. Dyke$^1$, K. Fenech$^1$, T. Peppler$^1$, M. G. Lingham$^1$, S. Hoinka$^1$, W. Zhang$^2$, S.-G. Peng$^3$, B. Mulkerin$^1$, H. Hu$^1$, X.-J. Liu$^1$, and C. J. Vale$^{1}$ \\

$^1$Centre for Quantum and Optical Sciences, Swinburne University of Technology, Melbourne 3122, Australia. \\ 
$^2$Department of Physics and Beijing Key Laboratory of Opto-electronic Functional Materials and Micro-nano Devices, Renmin University of China, Beijing 100872, China. \\ 
$^3$State Key Laboratory of of Magnetic Resonance and Atomic and Molecular Physics, Wuhan Institute of Physics and Mathematics, Chinese Academy of Sciences, Wuhan 430071, China. \\

\end{centering} 

\setcounter{figure}{0}
\renewcommand{\thefigure}{S\arabic{figure}} 

\section{Role of collisions during transverse expansion}

In this work we perform use of the transverse expansion rate of clouds released from a highly oblate 2D potential as a probe of the {\it in situ} kinematics. As the magnetic field is left on after release, the $s$-wave scattering length remains high and the expansion takes place in the collisional regime. Microscopically, this means that elastic collisions can occur during expansion which could redistribute energy from the radial motion into the transverse dimension, affecting the measured widths. It is therefore important to establish how significant this can be and what it means for the interpretation of our data. Here we consider a `worst-case scenario', to establish the largest effect that collisional dynamics could have on our data. By comparing this to our observations we readily see that the features in our data could not arise from collisions alone and must therefore originate from the {\it in situ} properties of the trapped cloud.

The expansion of both classical and quantum gases in the collisional regime can be described using a hydrodynamic picture \cite{skagan97,smenotti02}. A quantitative simulation of the experiments here lies beyond the scope of this work, however, in certain limiting cases, it is possible to answer basic questions about the expansion of gases in the collisional regime. Of relevance here, we can estimate how the energy transferred into the transverse dimension will scale with the total atom number $N$, starting from an initially 2D cloud.

Specifically, we consider the expansion of a 2D gas after removing the transverse confinement with the magnetic field tuned to the pole of a Feshbach resonance where the $s$-wave scattering length diverges and elastic collisions will be maximised. The amount of additional energy, $\Delta E_{(r \shortrightarrow z)}$, that can be transferred from the radial ($x$, $y$) dimensions to the transverse ($z$) dimension in the expansion time $\tau$ will be set by (see for example \cite{smonroe93})
\begin{equation}
\Delta E_{(r \shortrightarrow z)} \propto \int_0^{\tau} \Gamma_{el}(t) \bar{E}_{(r)} \mathrm{d}t,
\label{eq1}  
\end{equation}
where $\Gamma_{el}(t)$ is the elastic collision rate and $\bar{E}_{(r)}$ is the mean energy of particles in the radial dimension. At low temperature, $T < T_F$, $\bar{E}_{(r)}$ will be determined by the 2D Fermi energy, $\varepsilon_F = (\hbar^2 \pi/m) n_{2D}$, where $n_{2D}$ is the 2D density and $m$ is the atomic mass. In general, $\bar{E}_{(r)}$ may be time-dependent, however, as we are only concerned with very short expansion times, $\tau \ll \omega_r^{-1}$, we make the conservative assumption that $\bar{E}_{(r)}$ remains constant, i.e. that the radial kinetic energy is not significantly depleted via transfer into the $z$-dimension. 

Once the transverse confinement has been removed atoms will collide in a 3D environment. Moreover, as there now exists a continuum of transverse states available for particles to scatter into, Pauli suppression will not play a role for collisions which transfer energy into the $z$-dimension. Hence, we can use the classical expression for the elastic collision rate
\begin{equation}
\Gamma_{el}(t) = \langle n(t) \rangle \sigma_{el} \bar{v},
\label{eq2}  
\end{equation}
where the mean density is given by
\begin{equation}
\langle n(t) \rangle  = (1/N) \int_{-\infty}^{\infty} n(\mathbf{r},t)^2 \mathrm{d}^3\mathbf{r},
\label{eq3}  
\end{equation} 
$\sigma_{el}$ is the elastic collision cross-section and $\bar{v}$ is the mean collision velocity. At the pole of the Feshbach resonance $\sigma_{el} \rightarrow 4 \pi / \bar{k}^2$ where the mean collision wavevector $\bar{k}$ is simply proportional to $\bar{v}$. Eq.~(2) then immediately gives $\Gamma_{el}(t) \propto \langle n(t) \rangle / \bar{v}$. 

During the short time $\tau$ the only time dependence in $\langle n(t) \rangle$ will stem from expansion along $z$, as the radial density profile remains essentially static for times $t \ll \omega_r^{-1}$. To proceed, we make the conservative assumption that the time-dependence in $\langle n(t) \rangle$ is independent of the atom number, such that
\begin{equation}
\langle n(t) \rangle =  \langle n(0) \rangle .  f_z(t),
\label{eq4}  
\end{equation}
where $ \langle n(0) \rangle$ is the initial mean density and $f_z(t)$ is a time-dependent scaling function that describes evolution of the mean density along $z$. While not necessarily exact, this assumption allows us to find an upper limit on $\Delta E_{(r \shortrightarrow z)}$ due to collisions and also means we do not need to know the exact form of $f_z(t)$ as we are only ever concerned with its integrated value that will be the same for all $N$. The only term in Eq.~(4) that is sensitive to the atom number is $\langle n(0) \rangle$.

In an ideal gas we can evaluate $\langle n(t) \rangle$ analytically which, at $T = 0$, is proportional to $N^{1/2}$. Experimentally, we can determine $\langle n(0) \rangle$ for an interacting gas directly by integration over the 2D density profile according to Eq.~(3). From our data taken at the pole of the Feshbach resonance (832 G) we find that the mean 2D density scales as $\langle n(0) \rangle \propto N^{0.59 \pm 0.08}$ for atom numbers below $N_{2D}^{(Id.)}$, due to the effect of interactions.

With this we arrive at a simple prediction for how the energy transferred from the radial to the transverse dimension scales with the atom number.  As $\bar{E}_{(r)}$ (set by $\varepsilon_F$) is proportional to the density, 
\begin{equation}
\begin{split}
\Delta E_{(r \shortrightarrow z)} & \propto \int_0^{\tau} \Gamma_{el}(t) \bar{E}_{(r)} \mathrm{d}t \\
& \propto \frac{ \langle n(0) \rangle^2}{\bar{v}} \int_0^{\tau} f_z(t) \mathrm{d}t \\
& \propto \langle n(0) \rangle^{3/2}
\label{eq4}  
\end{split}
\end{equation}
where we have used $\bar{v} \propto \langle n(0) \rangle^{1/2}$ since, within our experimental uncertainties, the radial density has the same form for all atom numbers in the range considered here.

Taking our measured scaling of $\langle n(0) \rangle$ with the atom number we find that the maximum amount of energy that can be transferred into the transverse expansion scales with the atom number to the power of $\approx 0.9$. Hence the growth of the mean cloud width $\langle z^2 \rangle^{1/2}$ goes with the square root of this energy such that we expect the strongest possible dependence to be $\langle z^2 \rangle^{1/2} \propto N^{0.45(6)}$. 

\begin{figure}
\includegraphics[width=0.45\textwidth]{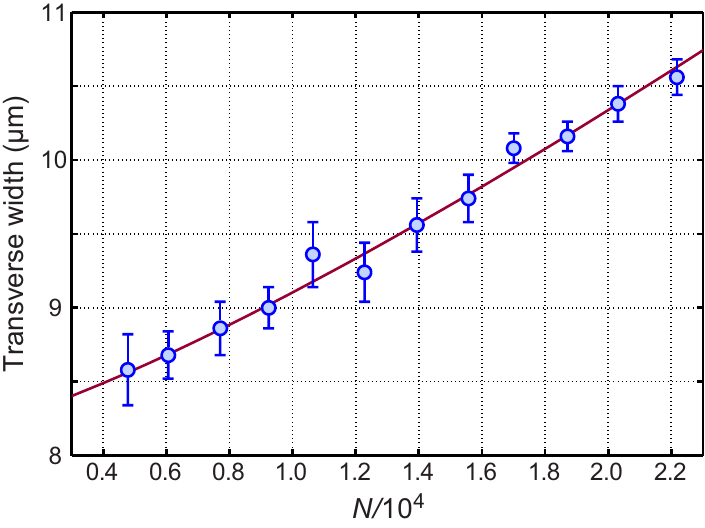}

\caption{The transverse width at $B_{0}=832.2$G for atom numbers $N < N_{2D}^{(Id.)}$. Power law fit to this data which shows that the width grows as $N^{1.24 \pm 0.16}$.}

\label{figs1}
\end{figure}

We can compare this prediction with data from Fig.~3(a) in the main paper. This is shown in Fig.~S1 above. Also plotted is a free power law fit to the data (blue line) which yields $\langle z^2 \rangle^{1/2} \propto N^{1.24 \pm 0.16}$. This is clearly a far stronger dependence on the atom number (approximately five standard deviations larger) than the predicted $N^{0.45}$ scaling from elastic collisions alone. We emphasise that, in arriving at this prediction, we assumed that the transverse width evolves independently of the atom number which we know from our data not to be the case. In reality we see that the density drops more rapidly for higher atom numbers which reduces the number of collisions during expansion such that the true $N$-dependence would be weaker than this. 

This analysis shows that, even though elastic collisions during expansion are expected to play a role, these alone can not account for the increased cloud widths we observe. Furthermore, while the effects of collisions will vary with the magnetic field, we expect that they will remain qualitatively similar on the BCS side of the Feshbach resonance. This power-law behavior is entirely inconsistent with the existence of a plateau for small atom numbers that we observe experimentally. We therefore conclude that our results provide clear evidence for the departure from 2D kinematics due to vibrationally excited atoms or molecules that were present in the trapped cloud.

\medskip

\section{Two-body expansion dynamics}

Here we present a theory for the expansion of the two-body ground state initially prepared in our two-dimensional trap, where the atoms are tightly confined in the transverse ($z$) direction, with weak radial confinement in the $x-y$ plane. We numerically calculate the two-body width $\sqrt{\left\langle z_{1}^{2}\right\rangle +\left\langle z_{2}^{2}\right\rangle }$ after the transverse confinement is removed and the cloud expands for time $t$ when the magnetic field is varied across the Feshbach resonance.

Here, we make the assumption that the atoms are confined in a purely harmonic potential, which means the center-of-mass (c.m.) motion becomes decoupled from the relative motion and we can treat the two components separately. 

\bigskip

\noindent \textbf{(i) Expansion of the center of mass}

\medskip

The c.m.~expansion represents a simple one-dimensional problem along the $z$-axis. At very low temperature $T$ ($k_{B}T\ll\hbar\omega_{z}$), where $\omega_{z}$ is the large transverse trap frequency, the two atoms should initially occupy the c.m. ground state $\varphi_{0}\left(Z\right)$ of the harmonic trap ($Z=\left(z_{1}+z_{2}\right)/2$ is the c.m.
coordinate). Once they are released the c.m. wavefunction at any time $t > 0$ takes the form
\begin{equation}
\Psi_{\text{c.m.}}\left(Z,t\right)=\int_{-\infty}^{\infty}C\left(K\right)\frac{e^{iKZ}}{\sqrt{2\pi}}e^{-i\hbar K^{2}t/4m}dK,\label{eq:cm1}
\end{equation}
where
\begin{equation}
C\left(K\right)=\int_{-\infty}^{\infty}\varphi_{0}\left(Z\right)\frac{e^{-iKZ}}{\sqrt{2\pi}}dZ=\left(\frac{d^{2}}{4\pi\eta}\right)^{1/4}e^{-d^{2}K^{2}/8\eta},\label{eq:cm2}
\end{equation}
$d=\sqrt{2\hbar/m\omega_{\rho}}$ , $\eta=\omega_{z}/\omega_{\rho}$ , $m$ is the atomic mass and $\omega_{\rho}$ is the radial trap frequency in the $x$-$y$ plane which remains on during the expansion. Inserting Eq.(\ref{eq:cm2}) into Eq.(\ref{eq:cm1}) yields
\begin{equation}
\Psi_{\text{c.m.}}\left(Z,t\right)=\left(\frac{4\eta}{\pi d^{2}}\right)^{1/4}\frac{e^{-2\eta Z^{2}/d^{2}\left(1+i\eta\omega_{\rho}t\right)}}{\sqrt{1+i\eta\omega_{\rho}t}}.\label{eq:cm3}
\end{equation}
Therefore, we easily obtain 
\begin{equation}
\left\langle Z^{2}\right\rangle =\int_{-\infty}^{\infty}\Psi_{\text{c.m.}}^{*}\left(Z,t\right)Z^{2}\Psi_{\text{c.m.}}\left(Z,t\right)dZ=\frac{d^{2}}{8\eta}\left(1+\eta^{2}\omega_{\rho}^{2}t\right).\label{eq:cm4}
\end{equation}

\bigskip

\noindent \textbf{(ii) Expansion of the relative motion}

\medskip

Initially, the atoms are trapped in a highly oblate three-dimensional harmonic potential ($\eta=\omega_{z}/\omega_{\rho}\approx200$). The Hamiltonian for the relative motion is 
\begin{equation}
\hat{H}_{i}=-\frac{\hbar^{2}}{m}\nabla^{2}+\frac{1}{4}m\omega_{\rho}^{2}\left(\rho^{2}+\eta^{2}z^{2}\right)+\hat{V}_{\text{int}}\left(\mathbf{r}\right),\label{eq:r1}
\end{equation}
where $\mathbf{r}\equiv\left(\boldsymbol{\rho},z\right)$ is the relative coordinate, and $\hat{V}_{\text{int}}\left(\mathbf{r}\right)$ is the interatomic interaction, which we simply treat as a contact interaction. By solving the Sch\"{o}rdinger equation for the relative-motion, the energy of the two-body bound state $E_{b}$ is determined by \cite{sIdziaszek2006A}
\begin{equation}
\frac{d}{a}=-\frac{1}{\sqrt{\pi}}\mathcal{F}_{i}\left(\epsilon_{0}\right),\label{eq:r2}
\end{equation}
where
\begin{equation}
\mathcal{F}_{i}\left(\epsilon_{0}\right)=\int_{0}^{\infty}\left[\frac{2\sqrt{\eta}e^{\epsilon_{0}\tau}}{\left(1-e^{-2\tau}\right)\sqrt{1-e^{-2\eta\tau}}}-\frac{1}{\sqrt{2}\tau^{3/2}}\right]d\tau,\label{eq:r3}
\end{equation}
$\epsilon_{0}=E_{b}/\hbar\omega_{\rho}-1-\eta/2<0$ , and $a$ is the $s$-wave scattering length. The corresponding bound-state wavefunction takes the form
\begin{equation}
\psi_{i}\left(\mathbf{r}\right)=\frac{C_{i}}{\sqrt{d}}\sum_{k=0}^{\infty}2^{\frac{2k-\epsilon_{0}}{2\eta}}\chi_{k}\left(\rho\right)\Gamma\left(\frac{2k-\epsilon_{0}}{2\eta}\right)D_{\frac{\epsilon_{0}-2k}{\eta}}\left(\sqrt{2\eta}\frac{\left|z\right|}{d}\right),\label{eq:r4}
\end{equation}
with the normalization coefficient
\begin{equation}
C_{i}=\left[\sum_{k=0}^{\infty}\frac{2\pi}{\sqrt{\eta}}\frac{\Gamma\left(\frac{2k-\epsilon_{0}}{2\eta}\right)}{\Gamma\left(\frac{2k-\epsilon_{0}}{2\eta}+\frac{1}{2}\right)}\beta\left(\frac{2k-\epsilon_{0}}{\eta}\right)\right]^{-1/2},\label{eq:r5}
\end{equation}
where 
\begin{equation}
\beta\left(x\right)\equiv\frac{1}{2}\left[\psi\left(\frac{1+x}{2}\right)-\psi\left(\frac{x}{2}\right)\right],\label{eq:r6}
\end{equation}
and $\psi\left(\cdot\right)$ is the digamma function. Here, $\Gamma\left(\cdot\right)$ is the gamma function, $D_{\nu}\left(\cdot\right)$ is the parabolic cylinder function, 
\begin{equation}
\chi_{k}\left(\rho\right)=\frac{e^{-\rho^{2}/2d^{2}}}{\sqrt{\pi}d}L_{k}\left(\frac{\rho^{2}}{d^{2}}\right),\label{eq:r7}
\end{equation}
 and $L_{k}\left(\cdot\right)$ is the Laguerre polynomial.

When the confinement along $z$ is switched off, the relative motion is described by the Hamiltonian 
\begin{equation}
\hat{H}=-\frac{\hbar^{2}}{m}\nabla^{2}+\frac{1}{4}m\omega_{\rho}^{2}\rho^{2}+\hat{V}_{\text{int}}\left(\mathbf{r}\right),\label{eq:r8}
\end{equation}
and the eigenvalue problem can be solved by following the route of Yurovsky and Olshanii \cite{sYurovsky2010R}. The eigenenergy $\epsilon=E/\hbar\omega_{\rho}-1$ is determined by
\begin{equation}
\frac{d}{a}=-\mathcal{F}\left(\epsilon\right),\label{eq:r9}
\end{equation}
 where
\begin{equation}
\mathcal{F}\left(\epsilon\right)=\sum_{n=0}^{\infty}\frac{-\cot\left(\frac{2L}{d}\sqrt{\epsilon/2-n}\right)-i}{\sqrt{\epsilon/2-n}}+\zeta\left(\frac{1}{2},-\frac{\epsilon}{2}\right),\label{eq:r10}
\end{equation}
$\zeta\left(\cdot,\cdot\right)$ is the Herwitz zeta function and $L$ is the length of the system along the $z$ axis. The corresponding wavefunction takes the form
\begin{equation}
\psi_{\epsilon}\left(\mathbf{r}\right)=\frac{C_{\epsilon}}{\sqrt{d}}\sum_{n=0}^{\infty}\chi_{n}\left(\rho\right)\frac{\cos\left[\frac{2L}{d}\sqrt{\epsilon/2-n}\left(\frac{z}{L}-1\right)\right]}{\sqrt{\epsilon/2-n}\sin\left(\frac{2L}{d}\sqrt{\epsilon/2-n}\right)},\label{eq:r11}
\end{equation}
with the normalization coefficient
\begin{equation}
C_{\epsilon}=\sqrt{2}\left\{ \sum_{n=0}^{\infty}\left[\frac{L}{d}\cdot\frac{1}{\left(\epsilon/2-n\right)\sin^{2}\left(\frac{2L}{d}\sqrt{\epsilon/2-n}\right)}+\frac{\cot\left(\frac{2L}{d}\sqrt{\epsilon/2-n}\right)}{2\left(\epsilon/2-n\right)^{3/2}}\right]\right\} ^{-1/2}.\label{eq:r12}
\end{equation}

From the derivation above, the transverse width of the relative-motion transverse after an expansion time $t$ can be evaluated numerically according to
\begin{equation}
\left\langle z^{2}\right\rangle =\sum_{\epsilon,\epsilon^{\prime}}A^{*}\left(\epsilon^{\prime}\right)A\left(\epsilon\right)\left\langle \psi_{\epsilon^{\prime}}\right|z^{2}\left|\psi_{\epsilon}\right\rangle e^{i\left(\epsilon-\epsilon^{\prime}\right)\omega_{\rho}t},\label{eq:r13}
\end{equation}
where
\begin{eqnarray}
A\left(\epsilon\right) & = & \int\psi_{\epsilon}^{*}\left(\mathbf{r}\right)\psi_{i}\left(\mathbf{r}\right)d^{3}\mathbf{r},\label{eq:r14}\\
\left\langle \psi_{\epsilon^{\prime}}\right|z^{2}\left|\psi_{\epsilon}\right\rangle  & = & \int\psi_{\epsilon^{\prime}}^{*}\left(\mathbf{r}\right)z^{2}\psi_{\epsilon}\left(\mathbf{r}\right)d^{3}\mathbf{r}.\label{eq:r15}
\end{eqnarray}
Finally, the total transverse width is easily obtained from Eqs.(\ref{eq:cm4}) and (\ref{eq:r13}),
\begin{equation}
\sqrt{\left\langle z_{1}^{2}\right\rangle +\left\langle z_{2}^{2}\right\rangle }=\sqrt{2\left\langle Z^{2}\right\rangle +\frac{1}{2}\left\langle z^{2}\right\rangle }.\label{eq:r16}
\end{equation}

\begin{figure}
\includegraphics[width=0.5\textwidth]{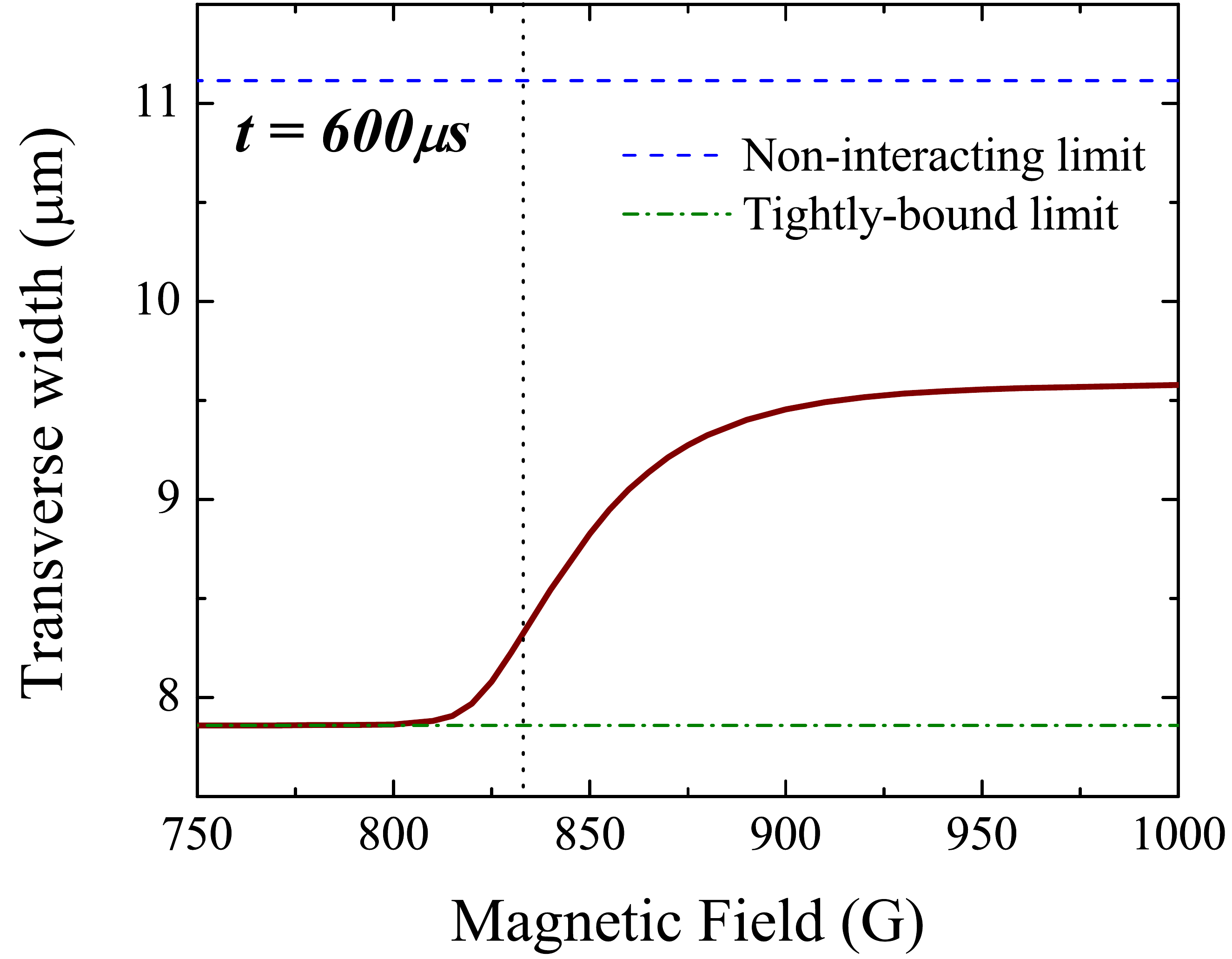}

\caption{The transverse width of the two-body state as a function of the magnetic field, after the expansion time $t=600\text{\ensuremath{\mu}s}$. In this calculation, the trap aspect ratio is $\eta=\omega_{z}/\omega_{\rho}=195$, and the length of the system along the $z$ axis is $L=10d$ . We include nearly $20000$ energy levels in the summation of Eq.~(\ref{eq:r13}), which shows good numerical convergence. The vertical dotted line indicates the position of the Feshbach resonance of $^{6}$Li centered at $B_{0}=832.2$G.}

\label{figs2}
\end{figure}

In Fig.~\ref{figs2} we plot the transverse width of the expanded two-body state as a function of the magnetic field across the Feshbach resonance. Note that to compare this with $\langle z^{2} \rangle^{1/2}$ considered in the experiments we need to divide Eq.~(25) by $\sqrt{2}$ which is based on the two-particle variance. The two-body width shrinks quickly to the tightly-bound limit on the BEC side of the resonance, where the main contribution is from the c.m.~part. Thus, far enough below the Feshbach resonance, the two-body width becomes insensitive to the magnetic field. Above the resonance the width grows but remains below the ideal atomic gas limit in this range of interactions. After release, the weak harmonic confinement in the $x$-$y$ plane remains on and is included in the calculations so the strict two-body ground state is very weakly bound even after removing the transverse confinement. The projection of the trapped two-body wavefunction onto the new basis includes significant overlap with this new ground state such that the transverse width after expansion remains well below the ideal gas case. In real experiments, the thermal energy will generally be higher than the ground state binding energy on the BCS side of the Feshbach resonance such that $k_B T \gg E_{b,3D}$ and pairs will readily break up in expansion. 

\end{document}